\documentclass[aps,preprint]{revtex4}
 \usepackage{epsfig}
\usepackage{amssymb}

\begin{document}

%%%%%%%%%%%%%%%%%%%%%%%%%%%%%%%%%% definitions

% \def\longvert{{\rule[-2mm]{0.1mm}{7mm}}\,}

\newcommand{\nc}{\newcommand}
\def \foot {\footnote}
\def\stroke{\vrule height8pt width0.4pt depth-0.1pt}
\def\topfleck{\vrule height8pt width0.5pt depth-5.9pt}
\def\botfleck{\vrule height2pt width0.5pt depth0.1pt}
\def\Zmath{\vcenter{\hbox{\numbers\rlap{\rlap{Z}\kern
0.8pt\topfleck}\kern 2.2pt\rlap Z\kern 6pt\botfleck\kern 1pt}}}
\def\Qmath{\vcenter{\hbox{\upright\rlap{\rlap{Q}\kern
3.8pt\stroke}\phantom{Q}}}}
\def\Nmath{\vcenter{\hbox{\upright\rlap{I}\kern 1.7pt N}}}
\def\Cmath{\vcenter{\hbox{\upright\rlap{\rlap{C}\kern
3.8pt\stroke}\phantom{C}}}}
\def\Rmath{\vcenter{\hbox{\upright\rlap{I}\kern 1.7pt R}}}
\def\Z{\ifmmode\Zmath\else$\Zmath$\fi}
\def\Q{\ifmmode\Qmath\else$\Qmath$\fi}
\def\N{\ifmmode\Nmath\else$\Nmath$\fi}
\def\C{\ifmmode\Cmath\else$\Cmath$\fi}
\def\R{\ifmmode\Rmath\else$\Rmath$\fi}

%%%%%%%%%%%%%%%%%%%%%%%%%%%%%%%%%%%%%%%%%%%%%%%%%%%%%%%%%%%%%%%%%%%%

\def\baselinestretch{1.0}

\begin{titlepage}

\vspace{1cm}

\begin{center}

{\Large \bf   Curves of Marginal Stability in Two-Dimensional
$CP(N-1)$ Models with $Z_N$-Symmetric Twisted Masses }
\vspace{0.5cm} \vspace{0.5cm} \vspace{0.5cm}

\end{center}

\begin{center}
{{\bf S. \"Olmez}$^{\,a}$, {\bf M.~Shifman}$^{\,b}$ }
\end {center}
$^a${\it Department of Physics and Astronomy, University of
Minnesota, Minneapolis, MN 55455, USA}\\
$^b${\it  William I. Fine Theoretical Physics Institute, University
of Minnesota, Minneapolis, MN 55455, USA}\\

\begin{center}
{\bf Abstract}
\end{center}
 We consider curves of marginal models in two dimensions with
$\mathcal{N} = (2, 2)$ supersymmetry. In these theories, one can
introduce twisted mass terms. The BPS spectrum has different number
of states in the weak and strong coupling regimes. This spectral
restructuring can be explained by the fact that two regimes are
separated by CMS on which some BPS states decay. We focus on a
special case of $Z_N$-symmetric twisted masses. In this case, the
general solution due to Dorey greatly simplifies, and CMS can be
found explicitly. For small-$N$ values numerical results are
presented. In the large-$N$ limit, we find CMS analytically up to
${\rm ln} N /N$ corrections.
\end{titlepage}
\section{Introduction}
 Two-dimensional $CP(N-1)$ sigma models with $\mathcal{N}
= (2, 2)$ supersymmetry present a rich theoretical laboratory. In
addition to the scale constant $\Lambda$, one can introduce other
dimensional parameters, the so-called twisted masses, which can be
interpreted as expectation values of a background twisted chiral
multiplet \cite{Alvarez et al,  Hanany Hori}. An exact description
of the spectrum of the Bogomol'nyi-Prasad-Sommerfield (BPS) states
as a function of the twisted masses is presented in Ref.
\cite{Dorey}. The spectrum of the theory with nonzero twisted masses
includes ``dyons'' in shortened multiplets. The dyon carries both
the topological and the Noether charges. The dyon mass is given by
the the absolute value of the sum of the topological mass, $m_{\cal
T}$, and the Noether mass, $m_{\cal N}$, which are complex
parameters,
\begin{equation}\label{Dyon mass}
    M=|m_{\cal T}+m_{\cal N}|.
\end{equation}
The triangular inequality for complex numbers gives $ M\leq|m_{\cal
T}|+|m_{\cal N}|$.  If the equality is satisfied,
\begin{equation}\label{Dyon mass satisfied}
    M=|m_{\cal T}|+|m_{\cal N}|,
\end{equation}
this is a boundary situation of a spectral restructuring. A
submanifold in the parameter space, where the equality is satisfied
is called CMS. If one crosses this manifold, discontinuities appear
in the spectrum. CMS, and the corresponding discontinuities of the
BPS spectrum, appear in theories with centrally extended
supersymmetry algebras \cite{vafa}. A detailed analysis of
metamorphosis of the BPS spectrum in the neighborhood of CMS is
given in Ref. \cite{Marginal Stab. and Metamorphism}. Dimension of
the submanifold, determined by the condition of the marginal
stability, need not be one. It can be larger depending on the number
of degrees of freedom residing in the twisted masses. First we must
note that the only condition on the twisted masses is
\begin{equation}\label{sum of the masses}
\sum_{l=0}^{N-1}m_l=0.
\end{equation}
This means that  we have $2(N-1)$ real independent parameters, for
arbitrary $N$. However, we will limit ourselves to a very special
and very interesting case of $Z_N$-symmetric masses,
\begin{equation}\label{Zn masses}
m_l= \,m\, e^{\frac{2 \pi i l}{N}},\;\;l=0,1,...,N-1.
\end{equation}
(Why it is of special physical interest is explained in Ref.
\cite{Gorsky Shifman Yung}.) If one introduces the masses $Z_N$
symmetrically, one has only two independent real parameters, which
come from the complex parameter $m$, for any value of $N$. It is
important to note that Eq. (\ref{sum of the masses}) is
automatically satisfied. The condition of CMS,  Eq. (\ref{Dyon mass
satisfied}), reduces the number of independent parameters on CMS
from two to one; thus, in this case CMS are indeed curves in the
complex $m$-plane. This is another reason to consider the theory
with $Z_N$-symmetric twisted masses. A general consideration of CMS
in the $CP(N-1$) model is presented in Ref. \cite{Dorey}. For $N=2$,
the explicit form of CMS is found in Ref. \cite{N=2 Solution}. In
this paper, we will consider $CP(N-1)$ sigma model with
$Z_N$-symmetric twisted masses and arbitrary $N$. We will find
explicit equations for CMS for any $N$. We give numerical solutions
for small $N$, and  show that for large $N$ CMS are circles with
\begin{equation}\label{cms in m plane}
    |m|=e \Lambda.
\end{equation}
The organization of the paper is as follows:\\ In Sect. \ref{Sect.
ZN symm.}, we introduce the $Z_N$-symmetric twisted masses and
derive the equation determining CMS. We solve this equation
numerically for $N=4$. In Sect. \ref{sect. Large N limit}, we
determine CMS explicitly for large values of $N$. The notation and a
brief introduction for our framework is given in Appendix. In the
first part of Appendix, we briefly introduce $N=2$ supersymmetric
field theories in two dimensions (for details see Ref. \cite{Witten
dimensional reduction}). In the second part, we discuss the central
extension of the theory and introduce the mirror representation.
Finally, we consider $CP(N-1)$ models with twisted masses and derive
the conditions on
the twisted masses to produce CMS.\\
\section{$Z_N$-Symmetric masses}\label{Sect. ZN symm.}
\begin{figure}[th] \centering
\includegraphics[width=2.5 in]{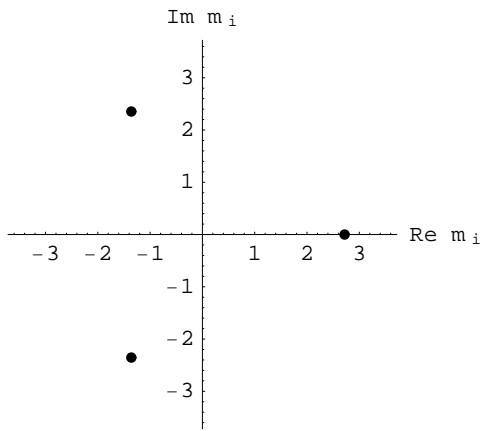}
\caption{$Z_3$-symmetric masses in the complex m plane}
\label{mass3}.
\end{figure}

\begin{figure}[th] \centering
\includegraphics[width=2.5 in]{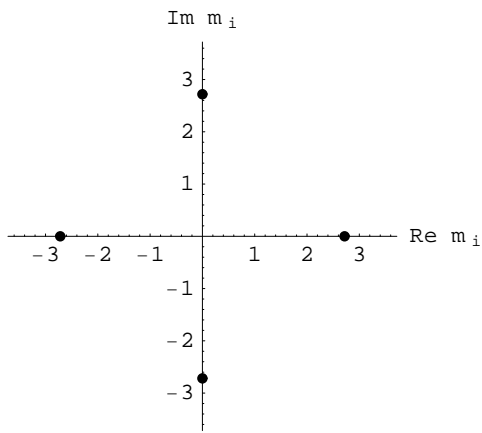}
\caption{$Z_4$-symmetric masses in the complex m plane}
\label{mass4}.
\end{figure}

 In the  $Z_N$-symmetric case, the form of the twisted masses is given in Eq. (\ref{Zn masses}), see Figs .
\ref{mass3} and \ref{mass4}, which show $Z_N$-symmetric masses for
$N=3$ and $N=4$, respectively.
The twisted masses are measured in the units of the scale constant
$\Lambda$ which is set to $1$ (see Appendix for details). Figures
\ref{mass3} and \ref{mass4} are plotted at $m=e$. In the general
case $m=\mu\, e^{i \theta}$ the scale of the corresponding plots
changes from $e$ to $\mu$ and they are rotated counterclockwise by
the angle $\theta$. \newpage
 The bosonic part of the $CP(N-1)$ model
deformed by twisted masses can be written as
\begin{equation}\label{bosonic lagrangian}
     S = \frac{2}{ g^2}\int d^2 x \left \{ |(\partial_{\alpha}- iA_{\alpha})
n^l|^2  + \sum_l |(\eta - m_l)n^l|^2 \right \}\,,
\end{equation}
with the condition $\sum_{l=0}^{N-1} \bar n^l n_l=1$ \cite{Gorsky
Shifman Yung}. $A_{\alpha}$ and $\eta$ are auxiliary fields, which
have no kinetic terms. $n^l$ ($l=0,\,1\,...,N-1$) is a complex
field. The second term in the action represents the twisted mass
deformation. Equation (\ref{bosonic lagrangian}) gives the action of
the $CP(N-1)$ sigma model in the linear gauged representation. The
fermionic part of the action can be constructed by requiring that
the action is ${\cal N}=2$ supersymmetric \cite{witten susy-sigma}.
The theory can be solved in $1/N$ expansion for large values of $N$
\cite{Large N expansion} (for nonvanishing twisted masses see Ref.
\cite{Gorsky Shifman Yung}). The theory has $N$ vacua, and for each
vacuum only one of the $n^l$ fields is nonvanishing. For example,
say, for the $k^{th}$ vacuum we can set $\eta=m_k$ so that the $l=k$
term in the sum vanishes. For the other terms to vanish we require
$n^l=0$ for $l\neq k$. If the twisted masses are of the form given
by Eq. (\ref{Zn masses}), then the action has an apparent $Z_N$
symmetry. This symmetry is spontaneously broken, as the vacuum is not $Z_N$ symmetric. \\
Although the representation given in Eq. (\ref{bosonic lagrangian})
is very transparent, it is not convenient for our purposes. It is
more convenient for us to work in the {\it mirror representation}
which is described in Appendix.\\
Hori and Vafa, who originally suggested the mirror representation,
derived it in the form of the Toda chain. Since then  a few other
equivalent representations were suggested. Following Dorey
\cite{Dorey} we will exploit a twisted chiral superfield $\Sigma$
representation.

\begin{figure}[th] \centering
\includegraphics[width=2.5 in]{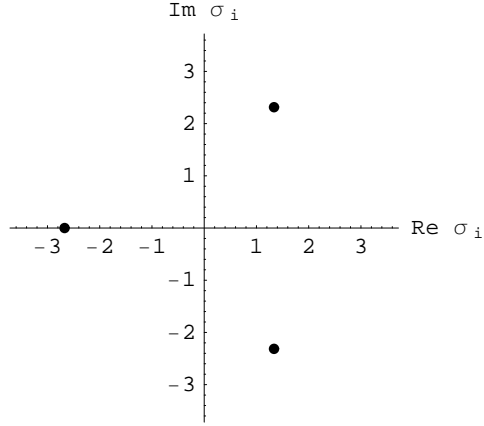}
\caption{$Z_3$-symmetric vacua in the complex $\sigma$-plane}
\label{vac3}.
\end{figure}

\begin{figure}[th] \centering
\includegraphics[width=2.5 in]{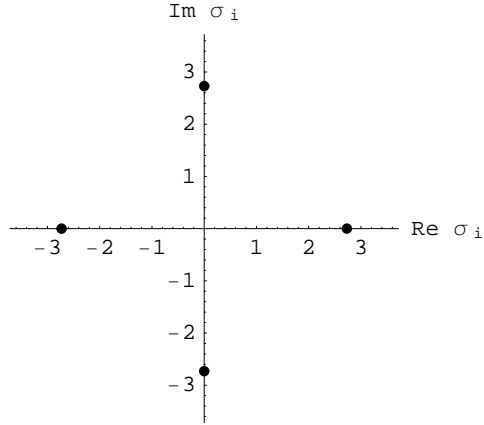}
\caption{$Z_4$-symmetric vacua in the complex $\sigma$-plane}
\label{vac4}.
\end{figure}

\begin{figure}[th] \centering
\includegraphics[width=2.5 in]{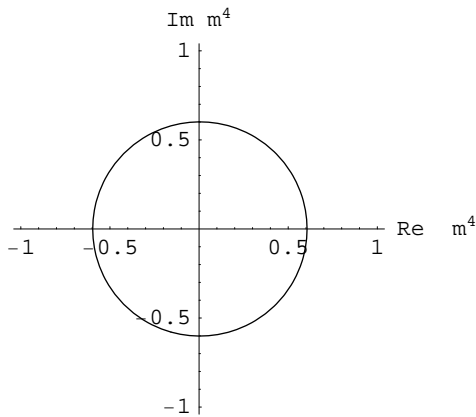}
\caption{The contour plot for CMS for $N=4$ $\;$ in the complex
$m^4$ plane}\label{N=4plot}
\end{figure}

In the mirror representation \cite{hori vafa}, the superpotential is
given by (see Appendix)
 \begin{equation}\label{superpotential with twisted masses 1}
   {\cal W}_{{\rm eff}}(\Sigma)=\frac{1}{4 \pi}\left( N \Sigma-\sum_{l=0}^{N-1}
m_l\, {\rm ln}\left(\frac{2}{\mu}(\Sigma+m_l)\right)\right),
 \end{equation}
 where $\Sigma$ is the twisted chiral field with the lowest
 component $\sigma$.
 The vacua of the theory are the solutions of
the following equation,
 \begin{equation}\label{vacua equation}
 \prod_{l=0}^{N-1}(\sigma+m_l)-1=0,
\end{equation}
where we set the scale constant $\Lambda=1$. The left-hand side of
this equation is a polynomial of degree $N$. In the general case, it
is not possible to find the roots of this equation analytically for
$N\geq 5$. However, the $Z_N$ symmetry of the twisted masses, given
in Eq. (\ref{Zn masses}), allows us to find the roots as,
\begin{equation}\label{roots as vacua}
    \sigma_k=\left(1+(-m)^N\right)^{\frac{1}{N} }e^{\frac{i 2 \pi
    k}{N}}.
\end{equation}
The vacua for $N=3$ and $N=4$ are shown in Figs. \ref{vac3} and
\ref{vac4}.
 Here we note the difference between the cases of  $odd$ and $even$
values of
 $N$.
 Figure \ref{vac3} is flipped with respect to Fig.
\ref{mass3}, but Fig. \ref{vac4} has the same form as Fig.
\ref{mass4}. Because of this difference, we will see that CMS will
be different for $odd$ and $even$ values of $N$ (at finite
$N$, not necessarily for large $N$).\\
 With the explicit solution for $\sigma_k$ given in Eq. (\ref{roots as vacua}), we can rewrite
${\cal W}_{{\rm eff}}$  in the critical points (see Eq.
(\ref{superpotential with twisted masses to be extremized}) and
(\ref{superpotential with twisted masses}) and Ref. \cite{Hanany
Hori,Dorey} ),
\begin{eqnarray}\label{vacuumfinal}
  {\cal W}_{{\rm eff}}(\sigma_k) &=& \frac{1}{4 \pi}\left( N
\sigma_k-\sum_{l=0}^{N-1} m  \,{\rm e}^{\frac{i 2 \pi l}{N}}{\rm
ln}\left(\sigma_k+m \,e^{\frac{i 2 \pi l}{N}}\right)\right)
\nonumber\\
   &=&\frac{1}{4 \pi}\,e^{\frac{i 2 \pi k}{N}}\left( N
\sigma_0-\,\sum_{l=0}^{N-1} m_l\, {\rm ln} \left(\sigma_0+m_l
\right)\right),
\end{eqnarray}
where \footnote{ Note that $\sigma_0$ is not well-defined at
$(-m)^N=-1$, at which $|m|=1$. However in the subsequent sections we
will see that $|m|=1$ is not on the CMS, so this will not affect our
discussion.}

\begin{equation}\label{vacuum definition}
    \sigma_0=\left(1+(-m)^N\right)^{\frac{1}{N}}.
\end{equation}
We used the fact that  $\sum_{l=0}^{N-1}m_l=0$, and also the angular
periodicity of the masses. Here we observe an important feature,
namely, the index $k$ in ${\cal W}_{{\rm eff}}(\sigma_k)$ appears
only in the phase and, as we will see, the phase factor will have no
impact on the CMS
consideration.
 Let us now consider a soliton interpolating between
 two vacua and carrying the topological charge $\overrightarrow{T}$. For each
allowed value of the topological charge $\overrightarrow{T}$, the
spectrum also includes an infinite tower of dyons with the global
charge $\overrightarrow{S}=s \overrightarrow{T}$, where $s\in Z$.
The vector $\overrightarrow{T}$ is of the form
$(0,...,-1,...,1,...,0)$ ( for instance, for a soliton interpolating
between the vacua $\sigma_k$ and $\sigma_l$,
$\overrightarrow{T}_k=-1$, and $\overrightarrow{T}_l=1$). One can
also introduce a topological mass vector,
\begin{equation}\label{Topological Mass Vector}
\overrightarrow{m_D}=({\cal W}_{{\rm eff}}(\sigma_0),{\cal W}_{{\rm
eff}}(\sigma_1),...,{\cal W}_{{\rm eff}}(\sigma_{N-1})).
\end{equation}
With these definitions, we can express the central charge in a
compact form,
\begin{equation}\label{Central Charge}
    Z=-i
(\overrightarrow{m}\cdot\overrightarrow{S}+\overrightarrow{m_D}\cdot\overrightarrow{T}).
\end{equation}
 The central charge, connecting the vacua $k$ and $l$, takes the
form
\begin{eqnarray}
Z_{k\,l}  &=&  -i(\overrightarrow{m}\cdot
    \overrightarrow{S_{k\,l}}+\overrightarrow{m_D}\cdot
    \overrightarrow{T_{k\,l}})
\nonumber\\
&=& -i\left(s(m_k-m_l)+(m_{Dk}-m_{Dl})\right)
\nonumber\\
 &=& -i\,m\,(e^{\frac{i 2 \pi k}{N}}-e^{\frac{i 2 \pi
l}{N}})\,\left\{s \,+\frac{2 \,i}{4 \pi m} \,\left(N
\sigma_0-\,\sum_{j=0}^{N-1} m_j\, {\rm ln}(\sigma_0+m_j )\right)
\right\}.
\end{eqnarray}
The overall factor $-i\,m\,(e^{\frac{i 2 \pi k}{N}}-e^{\frac{i 2 \pi
l}{N}})$ plays no role in the determination of CMS. The condition
for CMS is that the terms in the braces must have the same phase so
that $|Z|=|m_{\cal N}|+|m_{\cal T}|$ is satisfied. It is clear that
$s$ is a real number. This implies that the second term must be real
too on CMS. This, in turn, implies
\begin{equation}\label{CMS Condition}
     {\rm Re}\left\{ \frac{1}{2 \pi m} \left(N \sigma_0-\,\sum_{j=0}^{N-1}
m_j\, {\rm ln}(\sigma_0+m_j )\right) \right\}=0.
\end{equation}
Eq. (\ref{CMS Condition}) is our basic relation determining CMS. It
can be solved analytically for large $N$. The solution will be
presented in Sect. \ref{sect. Large N limit} . Small-$N$ solutions
can be found numerically. For $N=2$ this was done  in Ref. \cite{N=2
Solution}. For $N=2$, Eq. (\ref{CMS Condition}) reduces to
\begin{equation}\label{N=2 CMS EQN}
    {\rm Re}\left\{
   {\rm ln}\frac{1+\sqrt{1+4/m^2}}{1-\sqrt{1+4/m^2}}-2\sqrt{1+4/m^2}\right\}=0.
\end{equation}
\begin{figure}[th] \centering
\includegraphics[width=2.5 in]{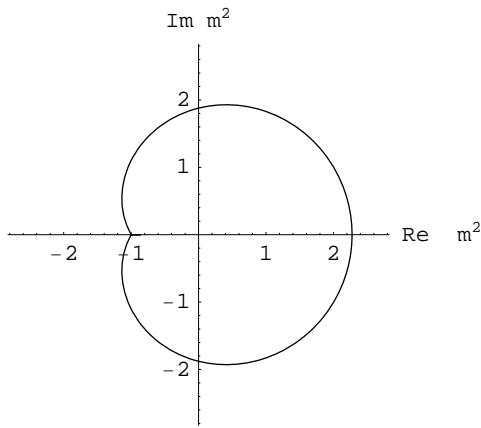}
\caption{The contour plot for CMS for $N=2$ $\;$ in the complex
$m^2$ plane}\label{N=2plot}
\end{figure}\\
The numerical solution is reproduced in Fig. \ref{N=2plot}. In Eq.
(\ref{N=2 CMS EQN}), we observe that the twisted mass parameter
appears in the form $m^2$, not $m$. For $N=2$ the physical sheet of
the Riemann surface is the complex $m^2$-plane, or, equivalently
half of the complex $m$-plane. We will see that this is a general
result; for generic $N$ the physical parameter is $m^N$ rather than
$m$ and, therefore, it is sufficient to solve Eq. (\ref{CMS
Condition}) for $|{\rm Arg}(m)|< \frac{\pi}{N}$, which is mapped
onto the complete complex $m^N$-plane. To illustrate the behavior of
states near CMS, let us consider an elementary state $\{T=0,\,
s=1\}$ where $T$ and $s$ show the topological and Noether charges,
respectively. From Fig. \ref{N=2plot}, we see that CMS cuts the real
axis at about $2.3$. For ${\rm Re}\,m^2$ slightly larger than this
value, the state $\{T=0,\, s=1\}$ becomes a marginally bound state
of two fundamental solitons $\{T=1,\,s=0\}$ and $\{T=-1,\, s=1\}$.
If we cross CMS and move to ${\rm Re}\,m^2$ smaller than $2.3$, the
interaction becomes repulsive and all the tower of excited states
disappears \cite{Marginal Stab. and Metamorphism, N=2 Solution}. For
$N=4$ CMS is given in Fig. \ref{N=4plot}. (The figure is scaled by
plotting $e^{-4} m^4$ rather than $m^4$.) We see that already at
$N=4$ CMS is pretty close to a circle. It becomes perfectly circular
at $N \rightarrow \infty$.
\section{The Large-$N$ Limit}\label{sect. Large N limit}
In the remainder of the paper we will construct CMS for large $N$.
Before delving into a detailed analysis, let us qualitatively
discuss the behavior of the function in Eq. (\ref{CMS Condition}).
The first term is of order $N$ whereas the second term, which is a
sum, has oscillating terms. Although there are $N$ terms, the result
of the summation will be order of $|m|$ rather than of order of $N$
due to this oscillatory behavior. If the sum is to be of the order
$N$, the argument of the logarithm must be exponentially small in
$N$ for at least some terms. The main strategy will be investigating
the sum to get a term of order $N$, which can cancel the term $N
\sigma_0$ in Eq. (\ref{CMS Condition}). In our analysis, we will
constrain ourselves to the region where $|{\rm Arg}(m)|<
\frac{\pi}{N}$ in the complex $m$-plane. This is the region which is
mapped onto the complete complex plane when we use $m^N$ as our
parameter instead of $m$. Due to $(-1)^N$ term in Eq. (\ref{roots as
vacua}), we see that it will be convenient to carry out the analysis
for $even$ and $odd$ $N$ values separately.
\subsection{CMS for large and $even$ $N$}
The results in the previous section show that a soliton is on the
CMS if
\begin{equation}\label{CMS CONDITION WITH ZN MASSES}
  {\rm Re}\left\{ \frac{1}{m} \left(N \sigma_0-\,\sum_{j=0}^{N-1} m_j
{\rm ln}(\sigma_0+m_j )\right) \right\}=0.
\end{equation}
To find the solution to this equation we will have to use slightly
different expansions of $\sigma_0$ depending on whether $|m|>1$,
$|m|=1$ or $|m|<1$, which suggests separate analysis of the problem
in three regions.
\subsubsection{$|m|>1$}
In this case, the terms
$j=\frac{N-2}{2},\,\frac{N}{2},\,\frac{N+2}{2}$ dominate the sum.
(Actually we will find out that the central term, $j=\frac{N}{2}$,
is the most dominant.) As $|m|>1$ and $|{\rm Arg}(m)|<
\frac{\pi}{N}$, we can use the expansion
$(1+m^N)^{\frac{1}{N}}\simeq m+\frac{1}{N\,m^{N-1}}$ to get
\begin{eqnarray}\label{N even M>1}
 \sum_{j=0}^{N-1} m_j {\rm ln}(\sigma_0+m_j )
\simeq &-& m\,{\rm ln}(\frac{1}{N\,m^{N-2}})
\nonumber\\[3mm]
&-& m\,(1-\frac{i 4 \pi }{N})\,{\rm ln}(\frac{1}{N
\,m^{N-2}}-m\,\frac{4 \pi i}{N})
\nonumber\\
&-&\,(1+\frac{i 4 \pi }{N})\,{\rm
ln}(\frac{1}{N\,m^{N-2}}+m\,\frac{4 \pi i}{N}).
\end{eqnarray}
The first line is the contribution coming from the $j=\frac{N}{2}$
term, whereas the second and the third lines  come from the
$j=\frac{N\pm 2}{2}$ terms. The first line presents the dominant
term in the large-$N$ limit. If we  define the twisted mass
parameter $m$ in polar coordinates, $m=\mu \, e^{i \theta}$, where
$\mu= |m|$, we can separate the real and imaginary parts as follows:
 \begin{equation}\label{N even r>1}
     \sum_{j=0}^{N-1} m_j ln(\sigma_0+m_j )=-m \,N\,\left[-\frac{N-1}{N}\,{\rm ln} \mu+O\left(\frac{{\rm ln}
     N}{N}\right)\right].
\end{equation}
Here  $O(\frac{{\rm ln} N}{N})$  stands for terms of order
$\frac{{\rm ln} N}{N}$. Now, Eq. \ref{CMS CONDITION WITH ZN MASSES}
implies
\begin{equation}\label{even N to be simplified}
N \, \left(1- \frac{N-1}{N}{\rm ln}\mu \right)=0,
\end{equation}
which in turn entails
\begin{equation}\label{N even result}
   \mu\,\rightarrow \,e
\end{equation}
as $N\rightarrow\,\infty$ ($N$ even). As we have discussed in the
$N=2$ case, the relevant parameter is $m^N$ rather than $m$. The
solution described above, which is an arc of a circle of radius $e$
and angle $|\theta|<\frac{\pi}{N}$, is mapped onto the complete
circle of radius $e^N$ in the complex $m^N$-plane.
\subsubsection{$|m|=1$}
In this case we can use the expansion $(1+m^N)^{\frac{1}{N}}\simeq
1+\frac{m}{N}\,{\rm ln}\,2 $. We can again approximate the sum by
the dominant terms, but in this case the argument of the logarithm
is of the order of $1/N$; hence it is impossible to cancel the
leading term of order $N$ in Eq.~(\ref{CMS Condition}) with this
${\rm ln} N$ term. Therefore, $|m|=1$ is not on the CMS for large
$N$ ($N$ $even$).
\subsubsection{$|m|<1$}
In order to have a complete and careful analysis, we can subdivide
this region into two parts as $|m|\rightarrow\,1-\epsilon$
($0<\epsilon \ll1$ ) and $|m|\ll1$. The main difference between the
two regions is that for the former, we can approximate the sum with
the dominant terms, whereas for the latter one each term produces a
contribution of the same order, and we convert the sum to an
integral, which is applicable in the large-$N$ limit. In both cases
the expansion \,$(1+m^N)^{\frac{1}{N}}\simeq 1+\frac{m^N}{N}$\, is
applicable. Let us start with the first case,
$|m|\rightarrow\,1-\epsilon$
\begin{eqnarray}
\sum_{j=0}^{N-1} m_j {\rm ln}(\sigma_0+m_j ) &\simeq& -m\,(1-\frac{i
4\pi }{N}){\rm ln}\left(1+\frac{m^N}{N}-m(1-\frac{i 4
\pi}{N})\right)
\nonumber\\[3mm]
&-&m\,(1+\frac{i 4 \pi }{N}){\rm
ln}\left(1+\frac{m^N}{N}-m(1+\frac{i 4 \pi}{N})\right)
\nonumber\\[3mm]
&-&m\,{\rm ln}(1+\frac{m^N}{N}-m).
\end{eqnarray}
Again, inspecting the arguments of the logarithms, we note that we
 end up with terms of order  ${\rm ln} N$, which cannot cancel the leading
 term in  Eq. (\ref{CMS Condition}). Therefore, we conclude that $|m|\rightarrow\,1-\epsilon$ does not belong to CMS.
In the second case, $|m|\ll1$, we need to change our strategy. The
sum cannot be approximated by a few dominant terms, as they
contribute almost equally. So we convert the sum to a corresponding
integral,
\begin{eqnarray}
\sum_{j=0}^{N-1} m_j {\rm ln}(\sigma_0+m_j ) &\simeq& \frac{N }{2
\pi i}\int_0^{\frac{2 \pi (N-1)}{N}} i m \,{\rm ln}(\sigma_0+m\,
e^{i x}) d x
\nonumber\\
&\simeq&-m \,{\rm ln}(1+m).
 \end{eqnarray}
This sum is of the order of $\,m\,{\rm ln}(m+1)\simeq \,O(1)\,$,
which cannot cancel the leading term in  Eq. (\ref{CMS Condition}).
Therefore, we conclude that $|m|\ll1$ does not belong to CMS either.
Combining all the results above, we see that in the large-$N$ limit
(with $N$ $even$ ),
\begin{equation}\label{CMS for even N in m^N plane}
    \mu^N(\theta)=e^N,-\pi<\theta<\pi,
\end{equation}
which means that CMS are circles of radius $e^N$ in the complex
$m^N$-plane (in the complex $m$-plane we have CMS at $|m|=e\,
\Lambda$).
\subsection{CMS for large and $odd$ N}
The analysis for $odd$ $N$ is slightly different than  the $even$
$N$ case. For $odd$ $N$, we have $\sigma_0=(1-m^N)^{\frac{1}{N}}$.
The large-$N$ expansion will have an extra phase factor compared to
the $even$ $N$ case. We will present the analysis for $|m|>1$, which
will produce the CMS for $odd$ $N$.
\subsubsection{$|m|>1$}
\begin{figure}[th] \centering
\includegraphics[width=2.5 in]{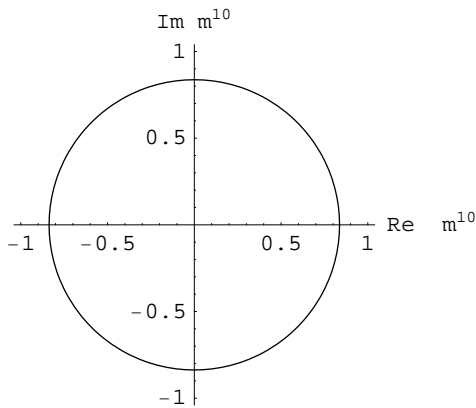}
\caption{The contour plot for CMS for $N=10$ $\;$in the complex
$m^{10}$ plane} \label{cms10}.
\end{figure}

\begin{figure}[th] \centering
\includegraphics[width=2.5 in]{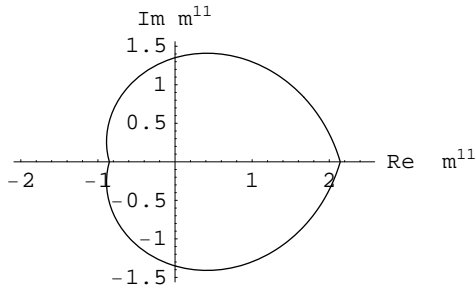}
\caption{The contour plot for CMS for $N=11$ $\;$in the complex
$m^{11}$ plane} \label{cms11}.
\end{figure}

 The expansion of $(1-m^N)^{\frac{1}{N}}$ depends on the phase of $m$.
For $- \frac{\pi}{N}<{\rm Arg}(m)<0$, we can use the expansion
$(1-m^N)^{\frac{1}{N}}\simeq
e^{i\frac{\pi}{N}}(m-\frac{1}{m^{N-1}}\frac{1}{N})$. In this case,
the $j=\frac{N+1}{2}$ term dominates the sum. With this expansion,
Eq. (\ref{CMS CONDITION WITH ZN MASSES}) reads,
\begin{equation}\label{odd N CMS 1}
{\rm Re}\{N e^{i \frac{\pi}{N}}+ e^{i \frac{\pi}{N}} {\rm ln}
(-m^{-N+1} e^{i \frac{\pi}{N}})\}=0,
\end{equation}
where we kept only $O(N)$ terms.   In the polar coordinates,
$m=\mu\, e^{i \theta}$ with $- \frac{\pi}{N}<\theta<0$, Eq.
(\ref{odd N CMS 1}) reduces to,
\begin{equation}\label{odd N CMS}
N -(1-N) {\rm ln} \mu + \frac{\pi}{N}(\pi+ N \theta) =0.
\end{equation}
This equation has the following solution,
\begin{equation}\label{Cms final for odd N negative phase}
\mu(\theta)=e^{1+\,\frac{\pi}{N^2}(\pi+ N \theta)}, \,\,\,\,\,\,\, -
\frac{\pi}{N}<\theta<0.
\end{equation}
At $N\rightarrow \infty$, this result reduces to the $even$-$N$
result, which is expected, of course. For $0<{\rm
Arg}(m)<\frac{\pi}{N}$, we can use the expansion
$(1-m^N)^{\frac{1}{N}}\simeq
e^{-i\frac{\pi}{N}}(m-\frac{1}{m^{N-1}}\frac{1}{N})$. Now the
$j=\frac{N-1}{2}$ term dominates the sum. Making the same steps we
get,
\begin{equation}\label{Cms final for odd N positive phase}
\mu(\theta)=e^{1+\,\frac{\pi}{N^2}(\pi- N \theta)}, \quad
0<\theta<\frac{\pi}{N}.
\end{equation}
We can combine both results as follows,
\begin{equation}\label{Cms final for odd N }
\mu(\theta)=e^{1+\,\frac{\pi}{N^2}(\pi- N |\theta|)}, \quad
-\frac{\pi}{N}<\theta<\frac{\pi}{N}.
\end{equation}
As in the $even$-$N$ case, we need to map this solution onto the
complex $m^N$-plane. Then CMS becomes,
\begin{equation}\label{CMS for odd N in m^N plane}
\mu^N(\theta)=e^{N+\,\frac{\pi}{N}(\pi- |\theta|)}, \quad
-\pi<\theta<\pi.
\end{equation}
This result reduces to the $even$-$N$ result at $N\rightarrow
\infty$. Collecting all the results for $even$ and $odd$ $N$, we
conclude that CMS are circles of radius $e^N$ in the complex
$m^N$-plane. Numerical solutions for CMS for $N=10$ and $N=11$ are
given in Figs. \ref{cms10} and \ref{cms11}.
We plotted $e^{-N} \,m^N$ rather than $m^N$ so that the radius
becomes unity in the large-$N$ limit. We note that  CMS for $N=10$
is a circle and its radius is slightly less than unity, which is the
case expected at $N\rightarrow \infty$. The deviation from circle is
more pronounced for $odd$ $N$, as seen in Fig. \ref{cms11}. This
behavior is consistent with the large-$N$ limit given in Eq. (
\ref{CMS for odd N in m^N plane}), from which we see that the radial
coordinate depends on the angle. At $\theta=0$, the radius in
enlarged by a factor of $e^{\frac{\pi^2}{N}}$, which is about $2.45$
at $N=11$. The $even-N$ result, Eq. ( \ref{even N to be
simplified}), gets an enlargement factor of $e^{\frac{1}{N}}$ for
any angle, which is close to unity at $N=10$. This explains why
$even-N$ results converge to $N\rightarrow\infty$ limit faster than
$odd-N$ results.
\section{Summary} In this paper, we discussed CMS in the
${\cal N}=(2,2)$ supersymmetric $CP(N-1)$ model with the
$Z_N$-symmetric twisted masses. The CMS condition is given by Eq.
(\ref{CMS Condition}). The solution to this equation is given in the
complex $m^N$-plane. We show that, for large values of $N$, CMS are
circles of radius of $e^N$ in the complex $m^N$-plane, which
corresponds to $|m|= e $. This result is approximate up to terms of
order $\frac{{\rm ln} N}{N}$ which die off rather slowly. This is
the reason to have different CMS curves for $odd$ and $even$ $N$,
for small $N$, as formulated in Eqs. (\ref{CMS for even N in m^N
plane}) and (\ref{CMS for odd N in m^N plane}) respectively. If we
consider the $N\rightarrow \infty$-limit, at which $\frac{{\rm ln}
N}{N}\ll1$, CMS are circles of radius $e^N$ for $odd$ and $even$
$N$. It is curious to note that in non-supersymmetric $CP(N-1)$
model the curve of the phase transition
is also circular at large $N$ \cite{Gorsky Shifman Yung}.\\
\section*{Acknowledgments}
The work of M.S. is supported in part by DOE grant DE-FG02-94ER408.

We are grateful to N. Dorey for useful discussions and comments.

\section{Appendix}
\renewcommand{\theequation}{A.\arabic{equation}}
\setcounter{equation}{0}
\subsection{$N=2$ Supersymmetry in Two Dimensions}
$N=1$  supersymmetry algebra \cite{wess bagger} has an $U(1)$
R-symmetry under which the left-handed supercharges have charge $-1$
and the right-handed ones have charge $+1$. One can obtain $N=2$
supersymmetry in two dimensions by dimensional reduction from $N=1$
supersymmetry in four dimensions \cite{Witten dimensional
reduction}. Eliminating the dependence of fields on two coordinates,
say $x^2$ and $x^3$, we get the two-dimensional Lorentz group and an
internal symmetry group associated with the rotations around the
eliminated coordinates. This internal symmetry is called the
$U(1)_{\rm A}$ symmetry. With this reduction, a left-handed spinor
in four dimensions becomes the Dirac spinor in two dimensions, which
consists of one left- and one right-handed spinor with the opposite
$U(1)_{\rm A}$ charges. The supercharges of the four-dimensional
theory  reduce to two Dirac spinors $Q_{L,R}$ and $\bar Q_{L,R}$.
$L,R$ shows the two-dimensional chirality whereas {\it bar} shows
the four-dimensional chirality. The Dirac spinors $Q_{L,R}$ and
$\bar Q_{L,R}$ carry the $U(1)_{\rm A}$ charges $-1,+1$ and $+1,-1$,
respectively. They are Hermitian conjugate to each other,
$(Q_{L,R})^\dag =\bar Q_{L,R}$. The anticommutation relations in two
dimensions can be written as
\begin{eqnarray}\label{commutations in 2 D}
  && \{Q_L,\bar Q_L\}\,\,=\,\,2(H+P),\nonumber \\
   &&\{Q_R,\bar Q_R\}\,\,=\,\,2(H-P),\nonumber \\
   &&Q_L^2=Q_R^2=\bar Q_L^2=\bar Q_R^2=0,
\end{eqnarray}
where $H$ and $P$ are the Hamiltonian and the momentum operators.
All the other commutators vanish unless there are central charges,
of which we will speak later. In two dimensions, the $U(1)$
R-symmetry of four-dimensional theory appears as another internal
symmetry which is called the $U(1)_{\rm V}$-symmetry. Under the
$U(1)_{\rm V}$ symmetry, the supercharges $Q_{L,R}$ and $\bar
Q_{L,R}$ have the charges $-1,-1$ and $+1,+1$, respectively. So, in
two dimensions, there are two $U(1)$ R-symmetry groups, $U(1)_{\rm
V}$ and $U(1)_{\rm A}$. The supercharges can be grouped as,
\begin{eqnarray}
Q_R\,\,&\,\,\bar Q_L\nonumber\\
Q_L\,\,&\,\,\bar Q_R
\end{eqnarray}
where the first (second) line has the $U(1)_{\rm A}$ charge $+1$
($-1$), and  the left (right) column  has the $U(1)_{\rm V}$ charge
$-1$ ($+1$). An important property of $N=2$ supersymmetry in two
dimensions is that it is possible to have a field $\Sigma$  which
obeys,
\begin{equation}\label{condition on twisted chiral multiplet in two dimensions}
\bar {\cal D}_L\Sigma=0= {\cal D}_R\Sigma,
\end{equation}
(compare with the chiral field which obeys $\bar {\cal
D}_L\Phi=0=\bar {\cal D}_R\Phi$) which is called the {\it twisted
chiral field }. Using the Bianchi identities, it is easy to get
$\Sigma$,
\begin{equation}\label{Twisted chiral field}
\Sigma=\frac{1}{2} \{\bar {\cal D}_L, {\cal D}_R\}.
\end{equation}
\subsection{Central Extension and Mirror Symmetry }
The $N=2$ supersymmetry algebra can be extended by the inclusion of
central charges which are associated with the topological charge of
the soliton sectors \cite{Witten olive susy with topological
charges}. As the central term should commute also with R-symmetry,
the central extension breaks the $U(1)_{\rm V}$ or/and the
$U(1)_{\rm A}$ symmetries. \\
For instance, consider a massive theory in which the $U(1)_{\rm V}$
symmetry is broken by a superpotential. Due to this central
extension, we have nonzero (anti)commutation relations in addition
to Eq. (\ref{commutations in 2 D}),
\begin{eqnarray}\label{commutations in 2 D with central extension}
\{Q_L,Q_R\}=2Z,\,\,&& \,\,\{\bar Q_L,\bar Q_R\}=2Z^*,
\nonumber\\
\{Q_L,\bar Q_R\}=0,\,\,&&\,\, \{\bar Q_L,Q_R\}=0,\nonumber\\
{[} F_{\rm A},Q_L {]} =-Q_L,~~ {[} F_{\rm A},Q_R {]} =Q_R,\!\!\!\!&&
{[} F_{\rm A},\bar Q_L {]} =\bar Q_L,~~ {[} F_{\rm A},\bar Q_R {]}
=-\bar Q_R,
\end{eqnarray} where $F_{\rm A}$ denotes the
generator of the $U(1)_{\rm A}$ R-symmetry. Using the
(anti)commutation relations in Eq. (\ref{commutations in 2 D}) and
(\ref{commutations in 2 D with central extension}) we observe that
the mass of the particle in a sector with central charge $Z$ is
bounded from below by
\begin{equation}\label{Central charge mass inequality}
   M\geq
|Z|\,.
\end{equation}
One can derive this result by calculating the anticommutator of the
operators $(H-P) Q_L- Z {\bar Q}_R$ and its Hermitian conjugate,
which is positive semi-definite by construction. The equality in Eq.
(\ref{Central charge mass inequality}) is satisfied if
\begin{equation}\label{BPS Bound}
(H-P)Q_L=Z\bar Q_R,
\end{equation}
which is called the {\it  BPS condition} . Consider eigenstates of
energy and momentum. For these eigenstates, Eq. (\ref{BPS Bound})
and its Hermitian conjugate imply that $Q_L$ and $\bar Q_L$ are
proportional to $\bar Q_R$ and $Q_R$ respectively. So the
supersymmetry multiplet is shortened. This is called a {\it BPS multiplet}. \\
We can also consider a theory in which the $U(1)_{\rm A}$ symmetry
is broken. In this case, in addition to Eq. (\ref{commutations in 2
D}), the algebra reads,
\begin{eqnarray}
\{Q_L,Q_R\}=0,\,\,&& \,\,\{\bar Q_L,\bar Q_R\}=0,
\nonumber\\
\{Q_L,\bar Q_R\}=2\widetilde{Z},\,\,&&\,\,
\{\bar Q_L,Q_R\}=2\widetilde{Z}^*,\nonumber\\
{[} F_{\rm V},Q_L {]} =-Q_L,~~ {[} F_{\rm V},Q_R {]}
=-Q_R,\!\!\!\!&& {[} F_{\rm V},\bar Q_L {]} =\bar Q_L,~~ {[} F_{\rm
V},\bar Q_R {]} =\bar Q_R,
\end{eqnarray} where $F_{\rm V}$ denotes the
generator of the $U(1)_{\rm V}$ R-symmetry. It is interesting to
note that the (anti)commutation relations would be the same in the
theories with broken $U(1)_{\rm V}$ and $U(1)_{\rm A}$ symmetry if
\begin{eqnarray}\label{mirror}
&&F_{\rm A}\longleftrightarrow F_{\rm V}\\[0.2cm]
&&Q_R\longleftrightarrow \bar Q_R.
\end{eqnarray}
This automorphism of the $N=2$ supersymmetry algebra is called the
{\it mirror symmetry} \cite{hori vafa}. \\
\subsection{CP(N-1) Models with Twisted
Masses} Consider a superrenormalizable $U(1)$ theory with $N$ chiral
superfields $\Phi_i$ with $+1$ charge, a gauge superfield and the
corresponding field strength $\Sigma$, which is a twisted chiral
superfield. The kinetic term and the interaction term is written as
a $D$-term in ${\cal N}=2$ superspace,
\begin{equation}\label{Dterm lagrangian}
{\cal L}_D=\int {\rm d}^4 \theta \left(\sum_{i=1}^N \bar\Phi_i e^{2
V}\Phi_i -\frac{1}{2 e^2} {\rm Tr } \bar\Sigma\Sigma\right).
\end{equation}
It is convenient to combine the Fayet-Iliopoulos term and the
topological $\theta$-term in the twisted $F$-term with the
Lagrangian,
\begin{equation}\label{Twisted F term lagrangian}
{\cal L}_F=\int {\rm d}^2 \theta {\cal W} (\Sigma)+ H.C. \, .
\end{equation}
 The twisted superpotential is

 \begin{equation}\label{Twisted superpotential}
{\cal W}(\Sigma)=\frac{i \tau \Sigma}{2},
\end{equation}
where $\tau= i r +\frac{\theta}{2 \pi}$ \cite{Witten dimensional
 reduction}.
 Now let us consider renormalizability of the theory without the
twisted masses. Gauge theories in two dimensions are
superrenormalizable. In our case, the only divergence comes from a
one-loop diagram and it can be absorbed into redefinition of the FI
parameter as follows,
\begin{equation}\label{renormalized FI parameter}
r(\mu)=r_0-\frac{N}{4\pi}{\rm ln}(\frac{M_{UV}^2}{\mu^2}),
\end{equation}
where $M_{UV}$ is the ultra violet cut-off and $\mu$ is the RG
subtraction scale. With the renormalized FI term, the superpotential
in Eq. (\ref{Twisted superpotential}) reads,
\begin{equation}\label{Renormalized
potential} {\cal W}_{{\rm eff}}(\Sigma)=\frac{i}{2}\Sigma \left(\hat
\tau-\frac{N}{2 \pi i}{\rm ln}(\frac{2 \Sigma}{\mu})\right),
\end{equation}
where $\hat\tau=i r(\mu)+\frac{\theta}{2\pi}+n^*$ with $n^*$ chosen
to minimize the potential energy. The condition for a supersymmetric
vacuum is,
\begin{equation}\label{condition for svac}
    \frac{\partial}{\partial\sigma}{{\cal W}}_{{\rm eff}}=0,
\end{equation}
where $\sigma$ is the lowest component of $\Sigma$. Eq.
(\ref{condition for svac}) has the following solution,
\begin{eqnarray}\label{sigma for susy vacua}
    \sigma^N&=&\left(\frac{\mu}{2}\right)^N e^{2\pi\, i\,
    \tau(\mu)} \equiv\tilde{\Lambda}^N,\nonumber\\
    \sigma_k&=&\tilde{\Lambda} e^{\frac{2\pi
    i k}{N}}
\end{eqnarray}
 The mass of the soliton interpolating between vacua $k$ and $l$ is
given by,
\begin{equation}\label{soliton mass}
    M=2|{\cal W}(\sigma_k)-{\cal W}(\sigma_l)|.
\end{equation}
When we include the twisted masses in our theory, we have to modify
Eq. (\ref{Renormalized potential}). Now the superpotential reads,
\begin{equation}\label{superpotential with twisted masses to be extremized}
 {\cal W}_{{\rm eff}}(\Sigma)=\frac{i}{2}\left(\Sigma
\hat \tau-\frac{1}{2 \pi i}\sum_{l=0}^{N-1} (\Sigma+m_l){\rm
ln}\left(\frac{2}{\mu}(\Sigma+m_l)\right)\right).
\end{equation}
$\hat\tau$ in this equation is determined by setting $
\frac{\partial}{\partial\sigma}
 W_{{\rm eff}}=0$. Imposing this condition for $\hat\tau$, we get,
 \begin{equation}\label{superpotential with twisted masses}
   {\cal W}_{{\rm eff}}(\Sigma)=\frac{1}{4 \pi}\left( N \Sigma-\sum_{l=0}^{N-1}
m_l\, {\rm ln}\left(\frac{2}{\mu}(\Sigma+m_l)\right)\right).
 \end{equation}
 This is the main formula that we will use to extract the topological
 masses of the solitons (for a pedagogical introduction to solitons see Ref. \cite{Rajaraman}). For this aim, we will also need the
 supersymmetric vacua as the solitons are the objects interpolating
 between different supersymmetric vacua. The equation for supersymmetric vacua is
 given by,
 \begin{equation}\label{vacua}
 \prod_{l=0}^{N-1}(\sigma+m_l)-\tilde\Lambda^N=0,
\end{equation}
which gives Eq. (\ref{sigma for susy
 vacua}) if the twisted masses are all vanishing. Calling the roots of this polynomial equation
$\sigma_l$, we see that there are $N$ supersymmetric vacua with
$\sigma=\sigma_l$. The BPS spectrum includes solitons interpolating
between different vacua, and carrying topological charges $T_i$ as
well as elementary particles carrying global $U(1)$ charges $S_i$.
For each pair of supersymmetric vacua, there exists a soliton
interpolating between them, which means that there are $\frac{N
(N-1)}{2}$ solitons carrying topological charge
$\overrightarrow{T}$. For each allowed value of the topological
charge $\overrightarrow{T}$, the spectrum also includes an infinite
tower of dyons with global charge $\overrightarrow{S}=s
\overrightarrow{T}$, where $s\in Z$. One can also introduce
topological mass vector,
\begin{equation}\label{Topological Mass Vector}
\overrightarrow{m_D}=({\cal W}_{{\rm eff}}(\sigma_0),{\cal W}_{{\rm
eff}}(\sigma_1),...,{\cal W}_{{\rm eff}}(\sigma_{N-1})).
\end{equation}
With these definitions, we can express the central charge as
\begin{equation}\label{Central Charge}
    Z=-i
(\overrightarrow{m}\cdot\overrightarrow{S}+\overrightarrow{m_D}\cdot\overrightarrow{T})
.
\end{equation}
In order to have a BPS state to decay into its constituents, its
mass must be equal to the sum of the masses of its constituents,
\begin{equation}\label{CMS cond}
   |Z|=|\overrightarrow{m}\cdot\overrightarrow{S}|+|\overrightarrow{m_D}\cdot\overrightarrow{T}|,
\end{equation}
 which is equivalent to requiring that each term in $Z$ to have the
same phase. This is the condition that determines CMS.

\end{document}